# Orthogonal, solenoidal, three-dimensional vector fields for no-slip boundary conditions

**Leaf Turner**
Department of Astronomy, Cornell University, Ithaca, NY 14853-6801, USA

Email: lt79@cornell.edu

**Abstract**

Viscous fluid dynamical calculations require no-slip boundary conditions. Numerical calculations of turbulence, as well as theoretical turbulence closure techniques, often depend upon a spectral decomposition of the flow fields. However, such calculations have been limited to two-dimensional situations. Here we present a method that yields orthogonal decompositions of incompressible, three-dimensional flow fields and apply it to periodic cylindrical and spherical no-slip boundaries.

## 1. Introduction

We derive orthogonal decompositions of arbitrary incompressible; i.e., solenoidal, three-dimensional velocity fields satisfying no-slip boundary conditions for periodic cylindrical and spherical geometries. The no-slip condition requires that all components of the velocity field vanish at the boundaries. Numerical calculations of fluid turbulence as well as turbulent closures depend on such spectral decompositions. To date, such no-slip viscous calculations have been limited to two-dimensional dynamics.

A progenitor of this analysis is the development of the scalar Chandrasekhar-Reid functions (Chandrasekhar and Reid 1957) used by Chandrasekhar and his coworkers to calculate no-slip fluid dynamics in a variety of situations, such as Couette flow (Chandrasekhar 1961). These functions have been assimilated into a two-dimensional, orthogonal, presumably complete, solenoidal vector basis satisfying no-slip boundary conditions for the study of turbulent cylindrical flow by Montgomery and his collaborators (Li and Montgomery 1996, Li *et al.* 1997). Our decompositions allow the application of spectral analyses to three-dimensional fluid turbulence with no-slip boundary conditions as well as to three-dimensional magnetohydrodynamic (MHD) turbulence. The latter has application to planetary and stellar MHD dynamos and may have relevance to the understanding of transport of quantities such as angular momentum in accretion disks around stars and black holes.

In Sec. 2, we obtain a three-dimensional basis appropriate for the description of an incompressible no-slip flow within a periodic cylinder. In Sec. 3, we derive a basis appropriate for the description of such a flow within a no-slip spherical boundary.

The vector fields in these bases are solenoidal; i.e., divergence-free. Because they are divergence-free, they are expressible in terms of curls. Furthermore, the divergence-free property implies that they are functions of only two scalar fields. For each geometry, we write down two classes of vector fields, each dependent on a scalar function. Requiring that these vector fields each vanish at the boundary imposes boundary conditions on the scalar fields. We then consider the inner products of these vector fields. After integrations by parts, we obtain matrix elements of a Hermitian operators between the different scalar fields, whose off-diagonal components must vanish for our orthogonal field representation.

Let us represent this Hermitian operator by $H_0$ and the scalar fields by $\xi_i(\mathbf{r})$. Then $\int d^3 r\, \mathbf{v}_i^*(\mathbf{r}) \cdot \mathbf{v}_j(\mathbf{r}) = \int d^3 r\, \xi_i^*(\mathbf{r}) H_0 \xi_j(\mathbf{r})$. Note that this is always positive for $i=j$, except for the trivial case of a null velocity. We demonstrate here how to find $\xi_i$ such that $\int d^3 r\, \xi_i^*(\mathbf{r}) H_0 \xi_j(\mathbf{r})$ (which equals $\int d^3 r\, \xi_j(\mathbf{r}) H_0 \xi_i^*(\mathbf{r})$ by virtue of the operator's Hermiticity) vanishes for $i \neq j$. In our examples, we shall see that we can construct a new Hermitian operator $H_c$ such that the equation,

$$(H_c + \lambda_k H_0)\xi_k(\mathbf{r}) = 0,$$

with the required boundary conditions on $\xi_k(\mathbf{r})$, can be solved analytically. The solutions for bounded domains will yield discrete values of the eigenvalue, $\lambda_k$. By taking the scalar product of this equation with a second eigenfunction, $\xi_j(\mathbf{r})$, next interchanging $k$ with $j$ and taking the complex conjugate, and finally subtracting the results using the Hermitian nature of these operators, we obtain

$$\int d\,r \left[\xi_j^*(\mathbf{r})(H_c + \lambda_k H_0)\xi_k(\mathbf{r}) - \xi_k(\mathbf{r})(H_c + \lambda_j^* H_0)\xi_j^*(\mathbf{r})\right] = (\lambda_k - \lambda_j^*)\int d^3 r\, \xi_j^*(\mathbf{r}) H_0 \xi_k(\mathbf{r}) = 0,$$

Since the integral on the right is always positive when $j=k$, the eigenvalues must all be real. For non-degenerate eigenvalues; i.e., $\lambda_j \neq \lambda_k$, $\int d^3 r\, \xi_j^*(\mathbf{r}) H_0 \xi_k(\mathbf{r}) = \int d^3 r\, \mathbf{v}_j^*(\mathbf{r}) \cdot \mathbf{v}_k(\mathbf{r}) = 0$, so that the set of the components of the associated velocity fields, $\{\mathbf{v}_i(r)\}$, provide a solenoidal, orthogonal basis. Because of the Hermitian nature of the operators, we believe the resulting basis to be complete.

**2. Periodic cylindrical geometry**

*2.1. Introduction*

In this section, we develop the spectral decomposition of a solenoidal vector field for describing an arbitrary flow within an annulus having two no-slip cylindrical boundaries: an inner boundary at $r = r_1$ and an outer one at $r = r_2$. (The no-slip condition means merely that all three components of the vector field vanish at the boundaries.) We assume a periodic length L in the z-direction. The desired solenoidal vector field is then expressible as a sum of two classes of components, which we shall see are themselves mutually orthogonal

$$\mathbf{v}(\mathbf{r}) = \sum_{m,l=-\infty}^{\infty} \sum_n c_{t,nml} \mathbf{v}_{t,nml}(\mathbf{r}) + \sum_{m,l=-\infty}^{\infty} \sum_n c_{p,nml} \mathbf{v}_{p,nml}(\mathbf{r}),$$
$$\mathbf{v}_{t,nml}(\mathbf{r}) = \nabla \times \{\psi_{nml}(r) \exp[i(m\theta + k_l z)]\hat{\mathbf{z}}\}, \qquad (1)$$
$$\mathbf{v}_{p,nml}(\mathbf{r}) = \nabla \times \nabla \times \{\Phi_{nml}(r) \exp[i(m\theta + k_l z)]\hat{\mathbf{z}}\}.$$

where $\hat{\mathbf{z}}$ is the unit vector along the z-direction. The wave number satisfies, $k_i = 2\pi i/L$. Each value of $m$ and $l$ of $\mathbf{v}$ is associated with two scalar potentials, $\sum_n c_{t,nml}\psi_{nml}(r)$ and $\sum_n c_{p,nml}\Phi_{nml}(r)$, as one would expect for a solenoidal vector field. We consider these two components separately.

We determine sets of scalar functions, $\{\psi_{nml}\}$ and $\{\Phi_{nml}\}$, such that the associated vectors are orthogonal; i.e., $\int_D \mathbf{v}^*_{t,nml}(\mathbf{r}) \cdot \mathbf{v}_{t,n'm'l'}(\mathbf{r}) d^3r = 0$ and $\int_D \mathbf{v}^*_{p,nml}(\mathbf{r}) \cdot \mathbf{v}_{p,n'm'l'}(\mathbf{r}) d^3r = 0$, for $n \neq n'$, $m \neq m'$, $l \neq l'$ and where the integration domain $D$ is over the annular volume between $r=r_1$ and $r=r_2$ along one period, L, in z. The asterisk denotes complex conjugation. These integrals vanish trivially if $m \neq m'$ or $l \neq l'$, so that we assume that both azimuthal and axial numbers are equal to $m$ and to $l$, respectively.

*2.2. Case 1: $ml \neq 0$*

Since $\mathbf{v}_{t,nml}(r) = \left[\frac{im}{r}\psi_{nml}(r)\hat{\mathbf{r}} - \frac{d\psi_{nml}(r)}{dr}\hat{\theta}\right] \exp[i(m\theta + k_l z)]$, the vanishing of $\mathbf{v}_{t,nml}$ at the boundaries imposes the conditions that both $\psi_{nml}$ and $\frac{\partial \psi_{nml}}{\partial r}$ vanish at the two boundaries: $r=r_1$ and $r=r_2$. Note that we are using the nomenclature $\hat{\mathbf{r}}$ and $\hat{\theta}$ to represent the unit radial and azimuthal vectors. Our orthogonality condition becomes

$\int_{r_1}^{r_2} r\, dr \left[\frac{m^2}{r^2}\psi^*_{nml}(r)\psi_{n'ml}(r) + \frac{d\psi^*_{nml}(r)}{dr}\frac{d\psi_{n'ml}(r)}{dr}\right] = 0$ for off-diagonal components; i.e., for

$n \neq n'$. An integration by parts using the vanishing of $\psi_n$ at the boundaries allows this orthogonality condition to be expressed as

$$\int_{r_1}^{r_2} r\, dr\, \psi^*_{nml}(r) \nabla^2_\perp \psi_{n'ml}(r) = 0, \text{ for } n \neq n' \qquad (2)$$

where

$$\nabla^2_\perp \equiv \frac{1}{r}\frac{d}{dr}\left(r\frac{d}{dr}\right) - \frac{m^2}{r^2}, \qquad (3)$$

represents the Laplacian operator that operates in the plane normal to the axial direction.

A set of scalar solutions $\{\psi_{jml}(r)\}$ satisfying (2) may be found by observing the Hermitian nature of the operators, $\nabla^2_\perp$ and $(\nabla^2_\perp)^2$:

$$\int_{r_1}^{r_2} r\, dr\, \psi^*_{nml}(r)\nabla^2_\perp \psi_{n'ml}(r) = \int_{r_1}^{r_2} r\, dr\, \psi_{n'ml}(r)\nabla^2_\perp \psi^*_{nml}(r),$$

$$\int_{r_1}^{r_2} r\, dr\, \psi^*_{nml}(r)(\nabla^2_\perp)^2 \psi_{n'ml}(r) = \int_{r_1}^{r_2} r\, dr\, \psi_{n'ml}(r)(\nabla^2_\perp)^2 \psi^*_{nml}(r). \qquad (4)$$

The first Hermitian property follows from an integration by parts using the vanishing of the solutions at the cylindrical boundaries; the second Hermitian property follows from two successive integrations by parts using both the vanishing of the solutions and of their radial derivative at the cylindrical boundaries.

A set of $\psi_{nml}$'s that satisfy (2) are the solutions of the fourth-order ordinary differential equation

$$\left(\nabla_\perp^2\right)^2 \psi_{nml}(r) + \alpha_{nml}^2 \nabla_\perp^2 \psi_{nml}(r) = \nabla_\perp^2 \left[\nabla_\perp^2 \psi_{nml}(r) + \alpha_{nml}^2 \psi_{nml}(r)\right] = 0 \tag{5}$$

satisfying the boundary conditions that $\psi_{nml}$ and its first derivative vanish at both $r=r_1$ and $r=r_2$. The discussion in Sec. 1 demonstrates that the eigenvalues, $\{\alpha_{nml}^2\}$, are real and that the solutions, $\psi_{nml}(r)$ and $\psi_{n'ml}(r)$, associated with two non-degenerate eigenvalues, $\alpha_{nml}^2$ and $\alpha_{n'ml}^2$, satisfy (2). The general form of these solution is

$$\psi_{nml}(r) = J_m(\alpha_{nml} r) + c_{Y,nml} Y_m(\alpha_{nml} r) + c_{+,nml} r^m + c_{-,nml} r^{-m}, \tag{6}$$

where the four constants, $c_{Y,nml}$, $c_{+,nml}$, $c_{-,nml}$, and $\alpha_{nml}$ are determined using the four boundary conditions. The functions, $J_m$ and $Y_m$, are the Bessel functions of the first and second kind of order $m$.

We next turn to the second class of solutions, those yielding $\mathbf{v}_{p,nml}(\mathbf{r})$. Here we wish to determine a set of scalar functions, $\{\Phi_{nml}(r)\}$, such that the associated vectors are orthogonal; i.e., $\int_D \mathbf{v}_{p,nml}^*(\mathbf{r}) \cdot \mathbf{v}_{p,n'm'l'}(\mathbf{r}) d^3r = 0$, for $n \neq n', m \neq m', l \neq l'$. Again, this integral vanishes trivially if $m \neq m'$ or $l \neq l'$, so that we assume that both azimuthal and axial numbers are equal to $m$ and $l$, respectively.

Since $\mathbf{v}_{p,nml}(\mathbf{r}) = \left\{ i k_l \left[\frac{d\Phi_{nml}(r)}{dr}\hat{\mathbf{r}} + \frac{im}{r}\Phi_{nml}(r)\hat{\theta}\right] - \nabla_\perp^2 \Phi_{nml}(r)\hat{\mathbf{z}}\right\} \exp[i(m\theta + k_l z)]$, the vanishing of $\mathbf{v}_{p,nml}(r)$ at the two boundaries requires that $\Phi_{nml}(r)$, as well as its first two radial derivatives, $d\Phi_{nml}(r)/dr$ and $d^2\Phi_{nml}(r)/dr^2$, vanish there. We thus have six boundary conditions on $\Phi_{nml}(r)$.

Our orthogonality condition becomes

$$\int_{r_1}^{r_2} r\, dr \left\{ k_l^2 \left[\frac{d\Phi_{nml}^*(r)}{dr}\frac{d\Phi_{n'ml}(r)}{dr} + \frac{m^2}{r^2}\Phi_{nml}^*(r)\Phi_{n'ml}(r)\right] + \nabla_\perp^2 \Phi_{nml}^*(r) \nabla_\perp^2 \Phi_{n'ml}(r)\right\} = 0$$

for off-diagonal components; i.e., for $n \neq n'$. When we integrate the first term involving derivatives of $\Phi$ once by parts using the vanishing of $\Phi$ at the boundaries, and the last term involving derivatives twice by parts using the vanishing of $\Phi$ and its first radial derivative at the boundaries, we transform this orthogonality condition to

$$\int_{r_1}^{r_2} r\, dr\, \Phi_{nml}^*(r) \nabla^2 \nabla_\perp^2 \Phi_{n'ml}(r) = 0 \text{ for } n \neq n', \tag{7}$$

where

$$\nabla^2 = \nabla_\perp^2 - k_l^2$$

represents the complete Laplacian operator for a state whose azimuthal and axial periodicities are contained in the exponential dependence: $\exp[i(m\theta + k_l z)]$.

A set of scalar solutions $\{\Phi_{iml}(r)\}$ satisfying (7) may be found by first noting the Hermitian nature of the operators, $\nabla^2 \nabla_\perp^2$ and $\nabla^2 (\nabla_\perp^2)^2$:

$$\int_{r_1}^{r_2} r\, dr\, \Phi^*_{nml}(r) \nabla^2 \nabla_\perp^2 \Phi_{n'ml}(r) = \int_{r_1}^{r_2} r\, dr\, \Phi_{n'ml}(r) \nabla^2 \nabla_\perp^2 \Phi^*_{nml}(r),$$

$$\int_{r_1}^{r_2} r\, dr\, \Phi^*_{nml}(r) \nabla^2 (\nabla_\perp^2)^2 \Phi_{n'ml}(r) = \int_{r_1}^{r_2} r\, dr\, \Phi_{n'ml}(r) \nabla^2 (\nabla_\perp^2)^2 \Phi^*_{nml}(r). \tag{8}$$

The first Hermitian property follows from two successive integrations by parts using the vanishing of the solutions and their radial derivates at the cylindrical boundaries; the second Hermitian property follows from three successive integrations by parts using both the vanishing of the solutions and of their first and second radial derivatives at the cylindrical boundaries.

A set of $\Phi_{nml}$'s that satisfy (7) are the solutions of the sixth-order ordinary differential equation

$$\nabla^2 (\nabla_\perp^2)^2 \Phi_{nml}(r) + \beta_{nml}^2 \nabla^2 \nabla_\perp^2 \Phi_{nml}(r) = \nabla^2 \nabla_\perp^2 [\nabla_\perp^2 \Phi_{nml}(r) + \beta_{nml}^2 \Phi_{nml}(r)] = 0 \tag{9}$$

satisfying the boundary conditions that $\Phi_{nml}$ as well as its first and second derivatives vanish at both $r=r_1$ and $r=r_2$. Again, the eigenvalues, $\{\beta_{nml}^2\}$, are real and that the solutions, $\Phi_{nml}(r)$ and $\Phi_{n'ml}(r)$, associated with two non-degenerate eigenvalues, $\beta_{nml}^2$ and $\beta_{n'ml}^2$, satisfy (7). The general form of these solution is

$$\Phi_{nml}(r) = J_m(\beta_{nml}r) + d_{Y,nml} Y_m(\beta_{nml}r) + d_{I,nml} I_m(k_l r) + d_{K,nml} K_m(k_l r) + d_{+,nml} r^m + d_{-,nml} r^{-m}, \tag{10}$$

where the six constants, $d_{Y,nml}, d_{I,nml}, d_{K,nml}, d_{+,nml}, d_{-,nml}$ and $\beta_{nml}$ are determined using the six boundary conditions. The functions, $I_m$ and $K_m$, are the modified Bessel functions of the first and second kind of order $m$.

*2.3. Case II: Translationally symmetric case: $m \neq 0, l = 0$*

The solution for $\mathbf{v}_t(\mathbf{r})$ clearly remains the same. However, the derivation of $\mathbf{v}_p(\mathbf{r})$ needs to be modified. The equation for $\mathbf{v}_{p,nm0}(r)$ becomes simply $\mathbf{v}_{p,nm0}(\mathbf{r}) = -\nabla_\perp^2 \Phi_{nm0}(r) \exp(im\theta) \hat{\mathbf{z}}$. We require that $\Phi_{nm0}(r)$ be the solution of

$$\nabla_\perp^2 \Phi_{nm0}(r) + \beta_{nm0}^2 \Phi_{nm0}(r) = 0, \tag{11}$$

where $\Phi_{nm0}(r)$ vanishes at the boundaries, $r=r_1$ and $r=r_2$, guaranteeing that $\mathbf{v}_{p,nm0}(r)$ satisfies the no-slip boundary condition. The solution of (11) is

$$\Phi_{nm0}(r) = J_m(\beta_{nm0}r) + d_{Y,nm0} Y_m(\beta_{nm0}r), \tag{12}$$

where the constant $d_{Y,nm0}$ and the (real) eigenvalue $\beta_{nm0}^2$ are determined from the vanishing of $\phi_{nm0}$ at the two boundaries. Note that the associated velocity fields satisfy the orthogonality condition that for $n \neq n'$, $\int_{r_1}^{r_2} r\, dr\, \mathbf{v}_{p,nm0}^*(r) \cdot \mathbf{v}_{p,n'm0}(r) = 0$.

*2.4. Case III: Cylindrically symmetric case: $m = 0$, $l \neq 0$*

For this case,
$$\mathbf{v}_{t,n0l}(\mathbf{r}) = -\frac{d\psi_{n0l}(r)}{dr}\exp[i k_l z]\hat{\theta}.$$

The no-slip boundary condition demands that $d\psi_{n0l}(r)/dr$ vanish at both $r = r_1$ and $r = r_2$. These states will be orthogonal for $n \neq n'$ if

$$\int_{r_1}^{r_2} r\, dr\, \frac{d\psi_{n0l}^*(r)}{dr}\frac{d\psi_{n'0l}(r)}{dr} = -\int_{r_1}^{r_2} r\, dr\, \psi_{n0l}^*(r)\left\{\frac{1}{r}\frac{d}{dr}\left[r\frac{d\psi_{n'0l}(r)}{dr}\right]\right\} = 0.$$

The second integral is obtained by performing an integration by parts and implementing the no-slip boundary condition on the $\psi$'s. This orthogonality condition will clearly be satisfied by requiring that

$$\frac{1}{r}\frac{d}{dr}\left[r\frac{d\psi_{n0l}(r)}{dr}\right] + \alpha_{n0l}^2\psi_{n0l}(r) = 0, \tag{13}$$

where $d\psi_{n0l}(r)/dr$ vanishes at both $r = r_1$ and $r = r_2$. The solution of (13) is

$$\psi_{n0l}(r) = J_0(\alpha_{n0l}r) + c_{Y,n0l}Y_0(\alpha_{n0l}r), \tag{14}$$

where the constant $c_{Y,n0l}$ and the real eigenvalue $\alpha_{n0l}^2$ are determined from the conditions on the two boundaries.

For the cylindrically symmetric components of $\mathbf{v}_p$, we note that

$$\mathbf{v}_{p,n0l}(\mathbf{r}) = \left\{i k_l \chi_{nl}(r)\hat{\mathbf{r}} - \frac{1}{r}\frac{d[r\chi_{nl}(r)]}{dr}\hat{\mathbf{z}}\right\}\exp[i k_l z],$$

where we have defined $\chi_{nl}(r) \equiv d\Phi_{n0l}(r)/dr$. The no-slip boundary conditions then requires that both $\chi_{nl}$ and its radial derivative vanish at the two boundaries, $r = r_1$ and $r = r_2$. The nontrivial part of the orthogonality condition legislates that for $n \neq n'$,

$$\int_{r_1}^{r_2} r\, dr\, \mathbf{v}_{p,n0l}^*(\mathbf{r}) \cdot \mathbf{v}_{p,n'0l}(\mathbf{r}) = \int_{r_1}^{r_2} r\, dr\, k_l^2 \chi_{nl}^*(r)\chi_{n'l}(r) + \int_{r_1}^{r_2} r\, dr\, \left\{\frac{1}{r}\frac{d[r\chi_{nl}^*(r)]}{dr}\right\}\left\{\frac{1}{r}\frac{d[r\chi_{n'l}(r)]}{dr}\right\}$$

$$= \int_{r_1}^{r_2} r\, dr\, \chi_{nl}^*(r)\left(k_l^2\chi_{n'l}(r) - \frac{d}{dr}\left\{\frac{1}{r}\frac{d[r\chi_{n'l}(r)]}{dr}\right\}\right) = 0,$$

where we utilized the vanishing of $\chi_n$ at the boundaries to obtain the final integral through an integration by parts.

If we define a Laplacian differential operator, $\aleph_l$, by

$$\aleph_l \chi_{nl}(r) \equiv \frac{d}{dr}\left\{\frac{1}{r}\frac{d}{dr}[r\chi_{nl}(r)]\right\} - k_l^2[r\chi_{nl}(r)],$$

we note that this orthogonality condition will be satisfied for solutions, $\{\chi_{nl}(r)\}$, of

$$(\aleph_l^2 + \beta_{nl}^2 \aleph_l)\chi_{nl}(r) = \aleph_l(\aleph_l + \beta_{nl}^2)\chi_{nl}(r) = 0 \tag{15}$$

that satisfy the no-slip boundary condition. This orthogonality results from the Hermitian property of $\aleph_l$ and $\aleph_l^2$ in this Hilbert space; i.e.,

$$\int_{r_1}^{r_2} r\, dr\, \chi_{nl}^*(r)\aleph_l \chi_{n'l}(r) = \int_{r_1}^{r_2} r\, dr\, \chi_{n'l}(r)\aleph_l \chi_{nl}^*(r),$$

$$\int_{r_1}^{r_2} r\, dr\, \chi_{nl}^*(r)\aleph_l^2 \chi_{n'l}(r) = \int_{r_1}^{r_2} r\, dr\, \chi_{n'l}(r)\aleph_l^2 \chi_{nl}^*(r).$$

The first condition arises from the vanishing of $\chi_{nl}$ at the cylindrical boundaries, $r_1$ and $r_2$; the second follows from the vanishing of both $\chi_{nl}$ and its radial derivative at the two boundaries. The solution of (15) is:

$$\chi_{nl}(r) = J_1(\gamma_{nl} r) + d_{Y,nl} Y_1(\gamma_{nl} r) + d_{I,nl} I_1(k_l r) + d_{K,nl} K_1(k_l r) \; ; \; \gamma_{nl} \equiv (\beta_{nl}^2 - k_l^2)^{\frac{1}{2}}. \tag{16}$$

The three constants, $d_{Y,nl}$, $d_{I,nl}$, and $d_{K,nl}$ and $\gamma_{nl}$ [which is related to the real eigenvalue, $\beta_{nl}^2$ according to (16)] are determined from the four boundary conditions; i.e., the vanishing of both $\chi_{nl}$ and its radial derivative at the two boundaries.

*2.5. Case IV: Axially and cylindrically symmetric case: m=0 and l=0*

For this case, the solution for $\mathbf{v}_t(\mathbf{r})$ proceeds exactly as in the case, $m=0, l\neq 0$:

$$\begin{aligned}\mathbf{v}_{t,n00}(\mathbf{r}) &= \nabla \times \psi_{n00}(r)\hat{\mathbf{z}} \\ \psi_{n00}(r) &= J_0(\alpha_{n00} r) + c_{Y,n00} Y_0(\alpha_{n00} r);\end{aligned} \tag{17}$$

where the constants, $c_{Y,n00}$ and $\alpha_{n00}$, are determined from the no-slip condition that $d\psi_{n00}(r)/dr$ vanishes at the two boundaries, $r = r_1$ and $r = r_2$.

Here the solution for $\mathbf{v}_p(\mathbf{r})$ proceeds as in the case, $m\neq 0, l=0$. This yields the result,

$$\begin{aligned}\mathbf{v}_{p,n00}(\mathbf{r}) &= \nabla \times [\nabla \times \Phi_{n00}(r)\hat{\mathbf{z}}], \\ \Phi_n(r) &= J_0(\beta_n r) + d_{Y,n00} Y_0(\beta_{n00} r),\end{aligned} \tag{18}$$

where the constants $d_{Y,n00}$ and $\beta_{n00}$ are determined from the vanishing of $\Phi_{n00}$ at the two boundaries, which guarantees no-slip.

*2.6. Orthogonality of the two classes of fields*

We demonstrate here that $\int_D d^3r\, \mathbf{v}^*_{t,nml}(\mathbf{r})\cdot \mathbf{v}_{p,n'm'l'}(\mathbf{r})=0$. For the case $m = 0$, the proof is trivial. For $m \neq 0$, we note that $\psi$ always vanishes at the boundary and is periodic in $z$. Hence the required integral can be transformed using a vector identity:

$$\int_D d^3r\, \mathbf{v}^*_{t,nml}(\mathbf{r})\cdot \mathbf{v}_{p,n'm'l'}(\mathbf{r})$$
$$= \int_D d^3r\, \nabla\cdot\left[\psi^*_{nml}(\mathbf{r})\hat{\mathbf{z}}\right]\times\left\{\nabla\times\left[\nabla\times\Phi_{n'm'l'}(\mathbf{r})\hat{\mathbf{z}}\right]\right\} - \int_D d^3r\, \psi^*_{nml}(\mathbf{r})\hat{\mathbf{z}}\cdot \nabla\times\nabla^2\Phi_{n'm'l'}(\mathbf{r})\hat{\mathbf{z}},$$

where we have set

$$\Phi_{nml}(\mathbf{r}) \equiv \Phi_{nml}(r)\exp[i(m\theta + k_l z)] \text{ and } \psi_{nml}(\mathbf{r}) \equiv \psi_{nml}(r)\exp[i(m\theta + k_l z)].$$

The first integral on the right-hand side vanishes because of the boundary condition on $\psi$. The second integral on the right-hand side vanishes trivially.

## 3. Spherical geometry

We now derive an orthogonal expansion basis for solenoidal fields that vanish at the boundaries of the spherical shell, $\rho_1 \leq \rho \leq \rho_2$. A point will be labeled by its coordinates, $(\rho,\theta,\phi)$. The associated unit vectors are $\hat{\rho},\hat{\theta},$ and $\hat{\phi}$. As with (1), a general solenoidal vector can be expanded in a basis involving two scalar functions for each set of "quantum numbers," $l$ and $m$:

$$\mathbf{v}(\mathbf{r}) = \sum_{l=0}^{\infty}\sum_{m=-l}^{l}\sum_n c_{t,nlm}\mathbf{v}_{t,nml}(\mathbf{r}) + \sum_{l=0}^{\infty}\sum_{m=-l}^{l}\sum_n c_{p,nlm}\mathbf{v}_{p,nlm}(\mathbf{r}),$$
$$\mathbf{v}_{t,nlm}(\mathbf{r}) = \nabla\times\left[\psi_{nlm}(\rho)Y_{lm}(\theta,\phi)\hat{\rho}\right], \qquad (19)$$
$$\mathbf{v}_{p,nlm}(\mathbf{r}) = \nabla\times\left\{\nabla\times\left[\Phi_{nlm}(\rho)Y_{lm}(\theta,\phi)\hat{\rho}\right]\right\},$$

where the $\{Y_{lm}(\theta,\phi)\}$ are the standard spherical harmonics.

We first determine the set of scalar functions, $\{\psi_{nlm}(\rho)\}$, such that the associated vectors are orthogonal; i.e., $\int_{D_s}\mathbf{v}^*_{t,nlm}(\mathbf{r})\cdot\mathbf{v}_{t,n'l'm'}(\mathbf{r})d^3r = 0$, for $n\neq n'$, $l\neq l'$, $m\neq m'$, and where the integration domain $D_s$ is the volume of a spherical shell bounded by $\rho = \rho_1$ and $\rho = \rho_2$.

Since $\mathbf{v}_{t,nlm}(\mathbf{r}) = \psi_{nlm}(\rho)\left[\frac{1}{r\sin\theta}\frac{\partial Y_{lm}(\theta,\phi)}{\partial \phi}\hat{\theta} - \frac{1}{r}\frac{\partial Y_{lm}(\theta,\phi)}{\partial \theta}\hat{\phi}\right]$, the no-slip boundary condition that requires $\mathbf{v}_{t,nlm}$ to vanish at the boundary implies that $\psi_{nlm}(\rho)$ vanishes at $\rho = \rho_1$ and $\rho = \rho_2$. The orthogonality condition requires that

$$\int_{D_s} d^3r \, \nabla \times [\psi_{nlm}^*(\rho) Y_{lm}^*(\theta,\phi)\hat{\rho}] \cdot \nabla \times [\psi_{n'l'm'}(\rho) Y_{l'm'}(\theta,\phi)\hat{\rho}]$$

$$= \int_{D_s} d^3r \, \nabla \cdot ([\psi_{nlm}^*(\rho) Y_{lm}^*(\theta,\phi)\hat{\rho}] \times \{\nabla \times [\psi_{n'l'm'}(\rho) Y_{l'm'}'(\theta,\phi)\hat{\rho}]\})$$

$$+ \int_{D_s} d^3r \, \psi_{nlm}^*(\rho) Y_{lm}^*(\theta,\phi)\hat{\rho} \cdot \nabla \times \{\nabla \times [\psi_{n'l'm'}(\rho) Y_{l'm'}(\theta,\phi)\hat{\rho}]\} = 0,$$

for $n \neq n', l \neq l'$, and $m \neq m'$. The boundary condition on $\psi$ eliminates the first integral on the right-hand side of this equation. Using the definition that the Laplacian operating in the surface normal to the spherical radius is

$$\nabla_\perp^2 \equiv \frac{1}{\rho^2 \sin(\theta)} \frac{\partial}{\partial \theta}\left[\sin(\theta) \frac{\partial}{\partial \theta}\right] + \frac{1}{\rho^2 \sin^2(\theta)} \frac{\partial^2}{\partial \phi^2} \quad , \tag{20}$$

the last integral can be rewritten compactly to yield the required orthogonality condition:

$$-\int_{D_s} d^3r \, \psi_{nlm}^*(\rho) Y_{lm}^*(\theta,\phi) \psi_{n'l'm'}(\rho) \nabla_\perp^2 Y_{l'm'}(\theta,\phi)$$

$$= l(l+1) \int_{D_s} d^3r \, \frac{\psi_{nlm}^*(\rho) Y_{lm}^*(\theta,\phi) \psi_{n'l'm'}(\rho) Y_{l'm'}(\theta,\phi)}{\rho^2} = 0,$$

if $n \neq n'$, $l \neq l'$, and $m \neq m'$. The orthogonality of the spherical harmonics guarantee that latter two conditions will be satisfied. Observe that if we set $\Xi_{nlm}(\rho) \equiv \frac{\Psi_{nlm}(\rho)}{\rho}$, the necessary condition can be re-expressed simply as

$$\int_{D_s} d^3r \, \Xi_{nlm}^*(\rho) Y_{lm}^*(\theta,\phi) \Xi_{n'l'm'}(\rho) Y_{l'm'}'(\theta,\phi) = 0 \tag{21}$$

when $n \neq n'$, $l \neq l'$, and $m \neq m'$.

Noting that the condition on the boundaries implies that the complete Laplacian operator is Hermitian; i.e., that

$$\int_{D_s} d^3r \, \Xi_{nlm}^*(\rho) Y_{lm}^*(\theta,\phi) \nabla^2 [\Xi_{n'l'm'}(\rho) Y_{l'm'}(\theta,\phi)] = \int_{D_s} d^3r \,^2 \Xi_{n'l'm'}(\rho) Y_{l'm'}(\theta,\phi) \nabla^2 [\Xi_{nlm}^*(\rho) Y_{lm}^*(\theta,\phi)],$$

we are assured that solutions of the Helmholtz equation

$$(\nabla^2 + \alpha_{nlm}^2) \Xi_{nlm}(\rho) Y_{lm}(\theta,\phi) = 0, \tag{22}$$

under the condition that $\Xi_{nlm}(\rho)$ vanish at the boundaries, have real eigenvalues $\alpha_{nlm}^2$ whose associated solutions satisfy (21) when the eigenvalues are distinct. The solutions of (22), independent of $m$, are $\Xi_{nlm}(\rho) = j_l(\alpha_{nl}\rho) + c_{y,nl} \, y_l(\alpha_{nl}\rho)$, where $j_l$ and $y_l$ are the spherical Bessel functions of the first and second kinds. We thus conclude that

$$\psi_{nlm}(\rho) = \rho[j_l(\alpha_{nl}\rho) + c_{y,nl} \, y_l(\alpha_{nl}\rho)]$$

are the desired functions yielding orthogonal fields, $\mathbf{v}_{t,nlm}(\mathbf{r})$. The constants, $\alpha_{nl}$, and $c_{y,nl}$ are determined from the vanishing of $\psi_{nlm}$ on the two boundaries, $\rho = \rho_1$ and $\rho = \rho_2$.

We now obtain the solutions for $\mathbf{v}_{p,nlm}(\mathbf{r})$. For each $l$ and $m$ value, we must find a set of scalar functions, $\{\Phi_{nlm}(\mathbf{r})\}$, such that the associated vectors are orthogonal;

i.e, $\int_D \mathbf{v}^*_{p,nlm}(\mathbf{r}) \cdot \mathbf{v}_{p,n'l'm'}(\mathbf{r}) d^3r = 0$, for $n \neq n', l \neq l'$, and $m \neq m'$. We note first from (19) that

$$\mathbf{v}_{p,nlm}(\mathbf{r}) = \left( -\hat{\rho}\nabla^2_\perp + \frac{\hat{\theta}}{\rho}\frac{\partial^2}{\partial\rho\partial\theta} + \frac{\hat{\phi}}{\rho\sin(\theta)}\frac{\partial^2}{\partial\rho\partial\phi} \right)[\Phi_{nlm}(\rho)Y_{lm}(\theta,\phi)].$$ Thus the no-slip

boundary conditions that legislate the vanishing of $\mathbf{v}_{p,nml}(\mathbf{r})$ at the boundaries, $\rho = \rho_1$ and $\rho = \rho_2$, are that $\Phi_{nlm}(\rho)$ and $d\Phi_{nlm}(\rho)/d\rho$ must each vanish at the two boundaries. The required condition of orthogonality is that for $n \neq n', l \neq l'$, and $m \neq m'$,

$$\int_{D_s} d^3r \, \nabla \times \{\nabla \times [\Phi^*_{nlm}(\rho)Y^*_{lm}(\theta,\phi)]\hat{\rho}\} \cdot \nabla \times \{\nabla \times [\Phi_{n'l'm'}(\rho)Y_{l'm'}(\theta,\phi)]\hat{\rho}\}$$

$$= \int_{D_s} d^3r \left\{ \nabla^2_\perp[\Phi^*_{nlm}(\rho)Y^*_{lm}(\theta,\phi)]\nabla^2_\perp[\Phi_{n'l'm'}(\rho)Y_{l'm'}(\theta,\phi)] + \frac{1}{\rho^2}\frac{\partial^2[\Phi^*_{nlm}(\rho)Y^*_{lm}(\theta,\phi)]}{\partial\rho\partial\theta}\frac{\partial^2[\Phi_{n'l'm'}(\rho)Y_{l'm'}(\theta,\phi)]}{\partial\rho\partial\theta} \right.$$

$$\left. + \frac{1}{\rho^2\sin^2(\theta)}\frac{\partial^2[\Phi^*_{nlm}(\rho)Y^*_{lm}(\theta,\phi)]}{\partial\rho\partial\phi}\frac{\partial^2[\Phi_{n'l'm'}(\rho)Y_{l'm'}(\theta,\phi)]}{\partial\rho\partial\phi} \right\} = 0.$$

(23)

We now do a couple of integrations by parts (or we could equally use well-known vector identities) to transform each of these integrals.

We consider initially the first of the three integrals. We invoke the vanishing of the $\Phi$'s at the boundaries to obtain the vanishing of the transverse divergence integrals:

$$\int_{D_s} d^3r \, \nabla^2_\perp[\Phi^*_{nlm}(\rho)Y^*_{lm}(\theta,\phi)]\nabla^2_\perp[\Phi_{n'l'm'}(\rho)Y_{l'm'}(\theta,\phi)]$$

$$= \int_{D_s} d^3r \, \vec{\nabla}_\perp \cdot \{\vec{\nabla}_\perp[\Phi^*_{nlm}(\rho)Y^*_{lm}(\theta,\phi)]\nabla^2_\perp[\Phi_{n'l'm'}(\rho)Y_{l'm'}(\theta,\phi)]\}$$

$$- \int_{D_s} d^3r \, \vec{\nabla}_\perp[\Phi^*_{nlm}(\rho)Y^*_{lm}(\theta,\phi)] \cdot \vec{\nabla}_\perp\{\nabla^2_\perp[\Phi_{n'l'm'}(\rho)Y_{l'm'}(\theta,\phi)]\}$$

$$= -\int_{D_s} d^3r \, \vec{\nabla}_\perp \cdot \left( [\Phi^*_{nlm}(\rho)Y^*_{lm}(\theta,\phi)] \cdot \vec{\nabla}_\perp\{\nabla^2_\perp[\Phi_{n'l'm'}(\rho)Y_{l'm'}(\theta,\phi)]\} \right)$$

$$+ \int_{D_s} d^3r \, \Phi^*_{nlm}(\rho)Y^*_{lm}(\theta,\phi)(\nabla^2_\perp)^2[\Phi_{n'l'm'}(\rho)Y_{l'm'}(\theta,\phi)]$$

$$= \int_{D_s} d^3r \, \Phi^*_{nlm}(\rho)Y^*_{lm}(\theta,\phi)(\nabla^2_\perp)^2[\Phi_{n'l'm'}(\rho)Y_{l'm'}(\theta,\phi)].$$

The second of the integrals on the right-hand side of (23) also may be transformed by performing two integrations by parts and again using the vanishing of the $\Phi$'s at the boundaries:

$$\int_0^\pi d\theta \sin(\theta) \int_0^{2\pi} d\phi \int_{\rho_1}^{\rho_2} \rho^2 d\rho \left[ \frac{1}{\rho} \frac{d\Phi_{nlm}^*(\rho)}{d\rho} \frac{\partial Y_{lm}^*(\theta,\phi)}{\partial\theta} \right]\left[ \frac{1}{\rho} \frac{d\Phi_{n'l'm'}(\rho)}{d\rho} \frac{\partial Y_{l'm'}(\theta,\phi)}{\partial\theta} \right]$$

$$= \int_0^\pi d\theta \sin(\theta) \int_0^{2\pi} d\phi \int_{\rho_1}^{\rho_2} d\rho \frac{\partial}{\partial\rho}\left\{ \left[ \Phi_{nlm}^*(\rho) \frac{\partial Y_{lm}^*(\theta,\phi)}{\partial\theta} \right]\left[ \frac{d\Phi_{n'l'm'}(\rho)}{d\rho} \frac{\partial Y_{l'm'}(\theta,\phi)}{\partial\theta} \right] \right\}$$

$$- \int_0^\pi d\theta \sin(\theta) \int_0^{2\pi} d\phi \int_{\rho_1}^{\rho_2} d\rho \, \Phi_{nlm}^*(\rho) \frac{\partial Y_{lm}^*(\theta,\phi)}{\partial\theta} \left[ \frac{d^2\Phi_{n'l'm'}(\rho)}{d\rho^2} \frac{\partial Y_{l'm'}(\theta,\phi)}{\partial\theta} \right]$$

$$= -\int_0^\pi d\theta \int_0^{2\pi} d\phi \int_{\rho_1}^{\rho_2} d\rho \frac{\partial}{\partial\theta}\left\{ \left[ \Phi_{nlm}^*(\rho) \sin(\theta) Y_{lm}^*(\theta,\phi) \right]\left[ \frac{d^2\Phi_{n'l'm'}(\rho)}{d\rho^2} \frac{\partial Y_{l'm'}(\theta,\phi)}{\partial\theta} \right] \right\}$$

$$+ \int_0^\pi d\theta \int_0^{2\pi} d\phi \int_{\rho_1}^{\rho_2} d\rho \, \Phi_{nlm}^*(\rho) Y_{lm}^*(\theta,\phi) \frac{d^2\Phi_{n'l'm'}(\rho)}{d\rho^2} \frac{\partial}{\partial\theta}\left[ \sin(\theta) \frac{\partial Y_{l'm'}(\theta,\phi)}{\partial\theta} \right]$$

$$= \int_{D_s} \Phi_{nlm}^*(\rho) Y_{lm}^*(\theta,\phi) \frac{d^2\Phi_{n'l'm'}(\rho)}{d\rho^2} \frac{\partial}{\rho^2 \partial\theta}\left[ \sin(\theta) \frac{\partial Y_{l'm'}(\theta,\phi)}{\partial\theta} \right].$$

The third integral on the right-hand side can be similarly transformed:

$$\int_{D_s} d^3r \left\{ \frac{1}{\rho^2 \sin^2(\theta)} \frac{\partial^2[\Phi_{nlm}^*(\rho) Y_{lm}^*(\theta,\phi)]}{\partial\rho\,\partial\phi} \frac{\partial^2[\Phi_{n'l'm'}(\rho) Y_{l'm'}(\theta,\phi)]}{\partial\rho\,\partial\phi} \right\}$$

$$= \int_{D_s} d^3r \, \Phi_{nlm}^*(\rho) Y_{lm}^*(\theta,\phi) \frac{1}{\rho^2 \sin^2(\theta)} \frac{\partial^4[\Phi_{n'l'm'}(\rho) Y_{l'm'}(\theta,\phi)]}{\partial\rho^2\,\partial\phi^2}.$$

Thus (23) can be expressed more concisely as

$$\int_{D_s} d^3r \, \nabla \times \left\{ \nabla \times [\Phi_{nlm}^*(\rho) Y_{lm}^*(\theta,\phi) \hat\rho] \right\} \cdot \nabla \times \left\{ \nabla \times [\Phi_{n'l'm'}(\rho) Y_{l'm'}(\theta,\phi) \hat\rho] \right\}$$

$$= \int_{D_s} d^3r \, \Phi_{nlm}^*(\rho) Y_{lm}^*(\theta,\phi) \nabla_\perp^2 \left( \nabla_\perp^2 + \frac{\partial^2}{\partial\rho^2} \right) \Phi_{n'l'm'}(\rho) Y_{l'm'}(\theta,\phi)$$

We now define

$$\Phi_{nlm}(\rho) \equiv \rho \, \chi_{nlm}(\rho).$$

Then,

$$\nabla_\perp^2 \left( \nabla_\perp^2 + \frac{\partial^2}{\partial\rho^2} \right) \Phi_{nlm}(\rho) Y_{lm}(\theta,\phi) = \rho \nabla_\perp^2 \nabla^2 \chi_{nlm}(\rho) Y_{lm}(\theta,\phi) = -\frac{l(l+1)}{\rho} \nabla^2 \chi_{nlm}(\rho) Y_{lm}(\theta,\phi),$$

where the $\nabla^2$ is the complete three-dimensional Laplacian in spherical coordinates. As a result, we note that

$$\int_{D_s} d^3r \, \nabla \times \{\nabla \times [\Phi^*_{nlm}(\rho)Y^*_{lm}(\theta,\phi)]\} \cdot \nabla \times \{\nabla \times [\Phi_{n'l'm'}(\rho)Y_{l'm'}(\theta,\phi)]\}$$

$$= -l(l+1) \int_{D_s} d^3r \, \chi^*_{nlm}(\rho)Y^*_{lm}(\theta,\phi)\nabla^2[\chi_{n'l'm'}(\rho)Y_{l'm'}(\theta,\phi)].$$

Since we wish this integral to vanish when $n \ne n'$, $l \ne l'$, and $m \ne m'$, and since it already vanishes if either of the last two conditions are met, we can set the condition for the $\chi$'s (for $l > 0$) that

$$\int_{\rho_1}^{\rho_2} \rho^2 d\rho \, \chi^*_{nlm}(\rho) \left[\frac{1}{\rho^2}\frac{\partial}{\partial \rho}\left(\rho^2 \frac{\partial}{\partial \rho}\right) - \frac{l(l+1)}{\rho^2}\right]\chi_{n'lm}(\rho) \text{ must vanish for } n \ne n'.$$

We now define the operator

$$\tilde{\nabla}^2 \equiv \frac{1}{\rho^2}\frac{\partial}{\partial \rho}\left(\rho^2 \frac{\partial}{\partial \rho}\right) - \frac{l(l+1)}{\rho^2}$$ and note that the boundary conditions imply that the

$\chi$'s and their radial derivatives vanish at $\rho = \rho_1$ and $\rho = \rho_2$. The Hermiticity of $\tilde{\nabla}^2$ and $\left(\tilde{\nabla}^2\right)^2$ follows immediately:

$$\int_{\rho_1}^{\rho_2} \rho^2 d\rho \, \chi^*_{nlm}(\rho)\tilde{\nabla}^2 \chi_{n'lm}(\rho) = \int_{\rho_1}^{\rho_2} \rho^2 d\rho \, \chi_{n'lm}(\rho)\tilde{\nabla}^2 \chi^*_{nlm}(\rho),$$

$$\int_{\rho_1}^{\rho_2} \rho^2 d\rho \, \chi^*_{nlm}(\rho)\left(\tilde{\nabla}^2\right)^2 \chi_{n'lm}(\rho) = \int_{\rho_1}^{\rho_2} \rho^2 d\rho \, \chi_{n'lm}(\rho)\left(\tilde{\nabla}^2\right)^2 \chi^*_{nlm}(\rho).$$

As we have seen, such Hermitian conditions guarantee that the solutions of

$$\tilde{\nabla}^2\left(\tilde{\nabla}^2 + \beta_l^2\right)\chi_{nlm}(\rho) = 0$$

satisfy the desired orthogonality condition. Since $\Phi_{nlm}(\rho) = \rho \chi_{nlm}(\rho)$, we obtain finally that

$\Phi_{nlm}(\rho) = \rho[j_l(\beta_l \rho) + d_{y,l} y_l(\beta_l \rho)] + d_{+,l}\rho^{l+1} + d_{-,l}\rho^{-l}$, where the four constants, $d_{y,l}$, $d_{+,l}$, $d_{-,l}$, and $\beta_l$ are determined by imposing the four boundary conditions that the $\Phi$'s and their radial derivatives must vanish at both $\rho = \rho_1$ and $\rho = \rho_2$.

One readily can verify that again the two classes of fields are themselves orthogonal; i.e, $\int_D d^3r \, \mathbf{v}^*_{t,nlm}(\mathbf{r}) \cdot \mathbf{v}_{p,n'l'm'}(\mathbf{r}) = 0$. We convert this integral as follows:

$$\int_D d^3r \, \mathbf{v}^*_{t,nlm}(\mathbf{r}) \cdot \mathbf{v}_{p,n'l'm'}(\mathbf{r})$$

$$= \int_D d^3r \, \nabla \cdot [\psi^*_{nlm}(\mathbf{r})\hat{\rho}] \times \{\nabla \times [\nabla \times \Phi_{n'm'l'}(\mathbf{r})\hat{\rho}]\} - \int_D d^3r \, \psi^*_{nlm}(\mathbf{r})\hat{\rho} \cdot \nabla \times \nabla^2[\Phi_{n'l'm'}(\mathbf{r})\hat{\rho}],$$

where we have set $\Phi_{nlm}(\mathbf{r}) \equiv \Phi_{nlm}(\rho)Y_{lm}(\theta,\phi)$ and $\psi_{nlm}(\mathbf{r}) \equiv \psi_{nlm}(\rho)Y_{lm}(\theta,\phi)$. The first integral on the right-hand side vanishes because of the vanishing of $\psi_{nlm}$ at the boundary. The second integral on the right-hand side vanishes trivially.

## 4. Graphical Results

We now present some interesting graphical results for $\mathbf{v}_p$. For the case of the cylinder having a radius of 1/2, Fig. 1 presents an example of the radial structure for the three components of this vector for the case: $k=2/3$, $m=3$ at $\theta=z=0$. The number of nodes is governed by the eigenvalue. Here we have chosen the eigenvalue, $\beta=334.43$. The radial, azimuthal, and axial components are depicted in Figs. 1(a,b,c), respectively. Note the curious structure of $\mathbf{v}_{p,\theta}$ in Fig. 1b.

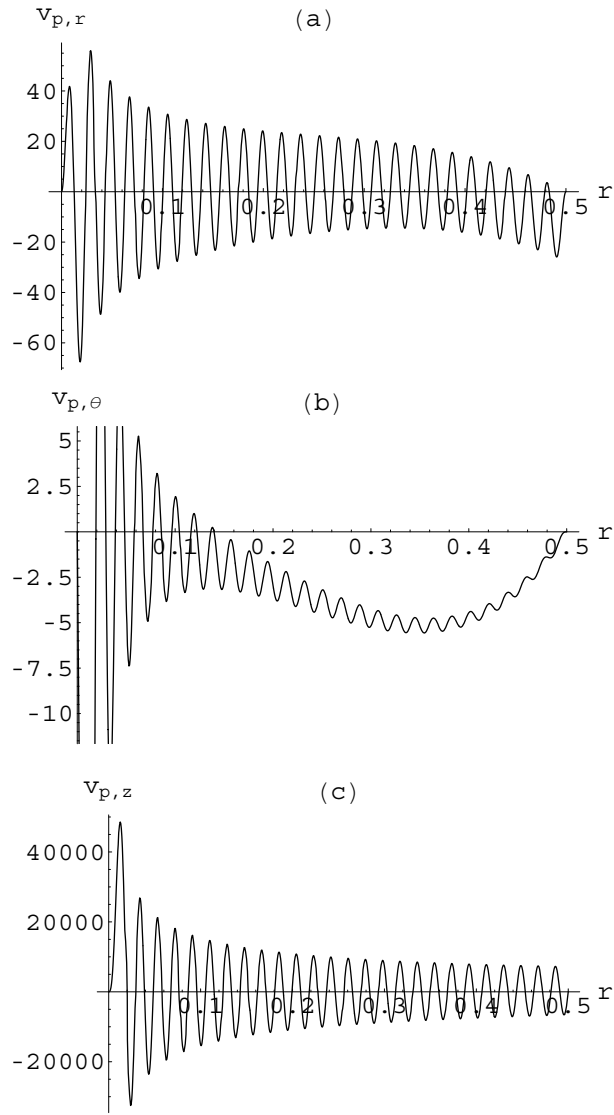

Figure 1

For the case of a sphere of unit radius, we present in Fig. 2(a,b,c) the three components of $\mathbf{v}_p$ for the case: $l=3$, $m=2$; namely, $\mathbf{v}_{p,\rho}, \mathbf{v}_{p,\theta}, \mathbf{v}_{p,\phi}$, respectively. The radial structure of $\mathbf{v}_p$ is shown at $\theta = 1$, $\phi = 0$. The number of nodes is again governed by the eigenvalue. Here we have chosen the eigenvalue, $\beta = 31.094$.

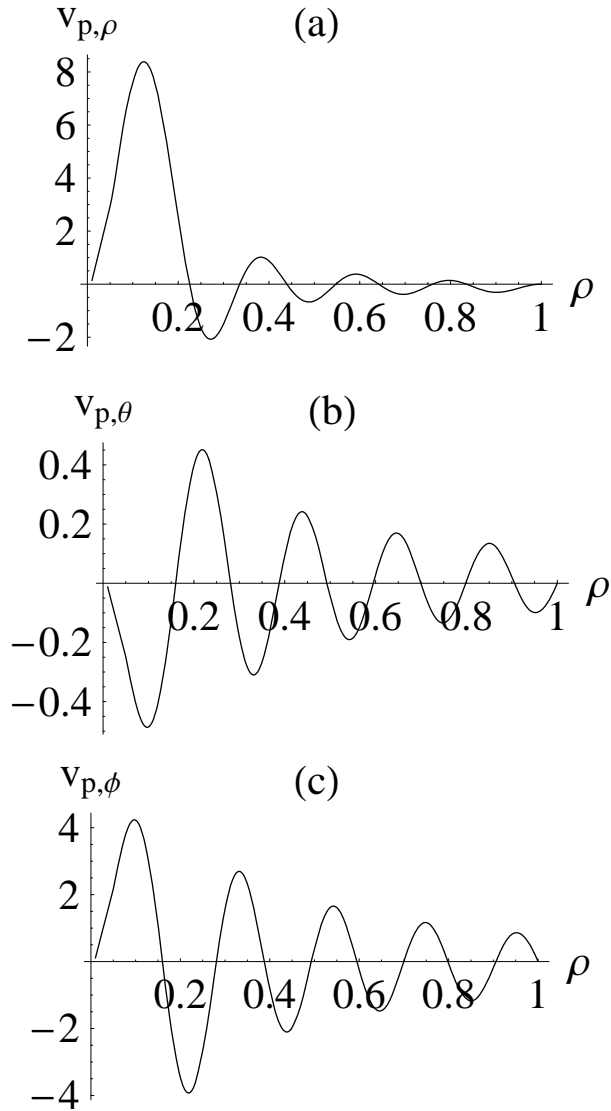

Figure 2

The ratio of the relative magnitudes of the components of each of these vectors at these high radial wave numbers is also noteworthy. Since these vectors are unnormalized, the actual magnitudes have no significance.

**5. Conclusion**

We have shown how to derive sets of orthogonal, solenoidal basis vectors that vanish on specified boundaries and have obtained these sets for both a periodic cylindrical boundary (including an cylindrical periodic annulus) and for a spherical boundary (including a spherical shell). We believe these sets of basis vectors provide a complete set for the expansion of an arbitrary solenoidal vector all of whose components vanish on the boundary. Our belief stems from the fact that they arise from solutions of differential equations that are self-adjoint; i.e., Hermitian, by construction. This method can be applied also to obtain the analogous basis for a slab geometry in which the vectors vanish on two infinite parallel planes.

We should emphasize that our construction is not necessarily unique. Merely as an example, note that in the second of the two equations of (8), the operator $\nabla^2\left(\nabla_\perp^2\right)^2$ could be replaced by $\left(\nabla_\perp^2\right)^3$. This yields a different differential equation that also provides orthogonal solutions, which however are not quite as elegant as those of (10).

**Acknowledgments**

We wish to thank Richard Lovelace for his encouragement of this research. We are grateful to David Montgomery for urging the expeditious preparation of this manuscript. Ari Turner and Richard Lovelace suggested improvements that I incorporated into the manuscript. Finally, we wish to thank Rena T. Fleur for her selfless dedication in painstakingly typing this manuscript. This work was supported in part by NSF grant AST-0507760.

**Figure Captions**

1. An example of a solenoidal velocity field satisfying no-slip boundary conditions within a cylinder of radius 1/2. The axial wave number, $k$, is 2/3; the azimuthal mode number, m, is 3. The radial profiles are presented at $\theta = z = 0$. The eigenvalue chosen is $\beta = 334.43$. Figs. 1a, 1b, and 1c, depict respectively the radial structures of $\mathbf{v}_{p,r}$, $\mathbf{v}_{p,\theta}$, and $\mathbf{v}_{p,z}$.

2. An example of a solenoidal velocity field satisfying no-slip boundary conditions within a sphere of unit radius. The values of *l* and *m* for this case are *3* and *2*, respectively. The radial profiles are presented at $\theta = 1$, $\phi = 0$. The eigenvalue chosen is $\beta = 31.094$. Figs. 2a, 2b, and 2c, depict respectively the radial structures of $\mathbf{v}_{p,\rho}$, $\mathbf{v}_{p,\theta}$, and $\mathbf{v}_{p,\phi}$.